\documentclass[a4paper,11pt]{article}
\usepackage{pos}

\title{Machine learning approaches for parameter reweighting in Monte-Carlo samples of top quark production in CMS}
\ShortTitle{Machine learning approaches for parameter reweighting in Monte-Carlo \\ samples of top quark production in CMS}
\author*[1]{Valentina Guglielmi}
\affiliation{Deutsches Elektronen-Synchrotron (DESY),\\
  Notkestrasse 85, Hamburg, Germany}
\note{For the CMS Collaboration.}
\emailAdd{valentina.guglielmi@desy.de}

\newcommand{\ttbar}{\textrm{\ensuremath{\mathrm{t\bar{t}}}}}
\newcommand{\mt}{\ensuremath{m_\text{t}}}
\newcommand{\pt}{\ensuremath{p_\text{{T}}}}
\newcommand{\hdamp}{\ensuremath{h_\text{{damp}}}}
\newcommand{\mtmc}{\ensuremath{m_\text{{t}}^\text{{MC}}}}

\abstract{
In high-energy particle physics, complex Monte Carlo (MC) simulations are needed to compare theory predictions to measurable quantities. 
Many and large MC samples are needed to be generated to take into account all the systematics.
Therefore, the MC statistics (and hence the MC modeling uncertainties) become a limiting factor for most measurements.
Moreover, the significant computational cost of these programs becomes a bottleneck in most physics analyses.
Therefore, it is extremely important to find a way to reduce the MC samples generated to decrease the MC statistical uncertainties and lower the computational cost. 
In these proceedings, we evaluate an approach called Deep neural network using Classification for Tuning and Reweighting (DCTR). 
DCTR is a method based on a Deep Neural Network (DNN) to reweight simulations to different models or model parameters and fit simulations, using the full kinematic information in the event. 
This reweighting methodology avoids the need for simulating the detector response multiple times by incorporating the relevant variations in a single sample.
In this way, the MC statistical uncertainties and the computational cost are both reduced.
Moreover, unlike the standard reweighting, in which the ratio in bins of two histograms at truth level
is performed, multidimensional and unbinned information can be used as inputs to the DNN.
In addition, DCTR can perform tasks that are not possible with other current existing methods, such as continuous reweighting as a function of any MC parameter, simultaneous reweighting of more MC parameters and tuning MC simulations to the data.
We test the method on MC simulations of top quark pair production, which we reweight to different SM parameter values and to different QCD models.

}

\FullConference{%
  41st International Conference on High Energy physics - ICHEP2022\\
  6-13 July, 2022\\
  Bologna, Italy
}

\begin{document}


\renewcommand{\logo}{\relax}
\maketitle

\section{Introduction}
In high-energy particle physics, complex Monte-Carlo (MC) simulations are needed to compare theory predictions to the measurable quantities, i.e. data. 
Many and large MC samples are needed to be generated to take into account all the systematics.
For this reason, MC modeling uncertainties are a limiting factor for most measurements. 
For example, in a recent measurement of the top quark-antiquark ($\ttbar$) pair production cross section in pp collisions at a centre of mass energy $\sqrt{s}=13$ TeV, 
where a simultaneous fit of the cross section of the $\ttbar$ system and of the MC mass of the quark top $\mtmc$ was performed~\cite{CMS:2018fks}, the main contribution to the $\mtmc$ was given by the MC statistics of the samples used to estimate the systematics.
Furthermore, the generation of many MC samples significantly increases the computational cost of these programs requiring to simulate the detector response multiple times.
A possible solution is to reweight the MC sample.
In this case, only the MC sample with the parameter nominal values is generated, while the variations are obtained by reweighting the nominal sample. 
Employing this strategy, only a sample is generated reducing MC statistics (hence MC modeling uncertainties) and lowering the computational cost.\\

Unlike the standard reweighting, in which the ratio in bins of two histograms at truth level is performed,
Deep neural network using Classification for Tuning and Reweighting (DCTR) is presented in these proceedings.
DCTR is an approach based on a Deep Neural Network (DNN) to reweight simulations to different models or model parameters using the full kinematic and flavor information in the event~\cite{Andreassen:2019nnm}.
While the standard reweighting is sensitive to the binning chosen and the dimension of the inputs could be 1, or at maximum 2, due to the increasing difficulty of the method with the number of dimensions, there are no restrictions on the size of the input feature space nor on the number of interpolated parameters using a DNN.
The first feature permits improving the reweighting precision using all the kinematic and flavor information of the event as inputs.
The second feature has as a consequence that simultaneous reweighting of more MC parameters can be performed, taking into account the correlations between the parameters.
Moreover, in addition to the advantages of the reweighting using a DNN, DCTR permits to perform several tasks not possible with the other methods.
First of all, DCTR is not limited to applying a discrete reweighting, but permits to perform a continuous reweighting as a function of any MC parameter.
In this way, a fast and more precise estimation of the systematic uncertainties is possible since any parameter value can be extrapolated.
Secondly, in addition to reweighting, the method can be also exploited to directly tune MC simulations to the data.
In this case, a classifier is used to construct the loss function used to optimize the simulation parameters.


\section{Deep neural networks using Classification for Tuning and
Reweighting}

\begin{figure}[htbp]
\centering
\includegraphics[width=0.9\textwidth]{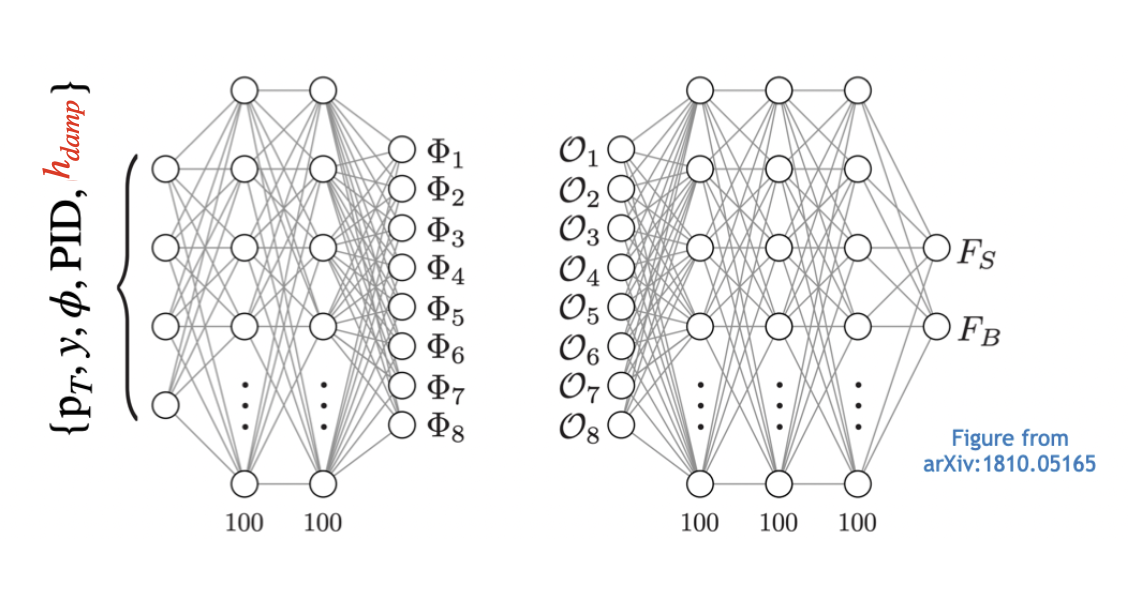}
\caption{The PFN architecture used in the DCTR method is depicted. 
It parametrizes the per-particle mapping $\Phi$ (on the left) and the function F (on the right), shown for the case of a latent space of dimension $l = 8$. 
The latent observable is $\ensuremath{O_\mathrm{a}} = \sum_{i}\ensuremath{\Phi_\mathrm{n}}(\pt, \ensuremath{y_\mathrm{i}}, \ensuremath{\phi_\mathrm{i}}, \ensuremath{m_\mathrm{i}}, \ensuremath{{pdgid}_\mathrm{i}})$~\cite{Komiske:2018cqr}.}
\label{fig:NN2}
\end{figure}

The default MC sample of top pair production in CMS is generated using the Heavy Quark Process (HVQ)~\cite{Frixione:2007nw} of the event generator Powheg~\cite{Frixione:2007vw}\cite{Alioli:2010xd}.
In Powheg, the resummation of the next-to-leading-order radiation is regulated by the $\hdamp$ variable, which enters the damping parameter D as in Eq.~\ref{eq:hdamp}:
\begin{equation}
    D = \dfrac{\hdamp^2}{\pt^{2} + \hdamp^2}
    \label{eq:hdamp}
\end{equation}
where $\pt$ is the transverse momentum of the particle and $\hdamp$ a parameter defined as $\hdamp = h\times \mt$, where $\mt$ is the mass of the quark top and $h$ is a real number.
This parameter affects the kinematic information of the entire event, such as the transverse momentum and the pseudorapidity of the $\ttbar$ system ($\pt(\ttbar)$, $\eta (\ttbar)$).
Since the parameter $\hdamp$ is not physical, an arbitrary value must be chosen in the simulation and varied to calculate the associated systematic uncertainty. As the parameter cannot be reweighted internally by MC generators, and $\hdamp$ variations are important in many top quark precision studies, this parameter is especially well suited to be implemented in the DCTR method.
Therefore, variations of $\hdamp$ in $\ttbar$ system were studied to test how well the method works.\\ 
\begin{figure}[htbp]
\centering
\includegraphics[width=0.50\textwidth]{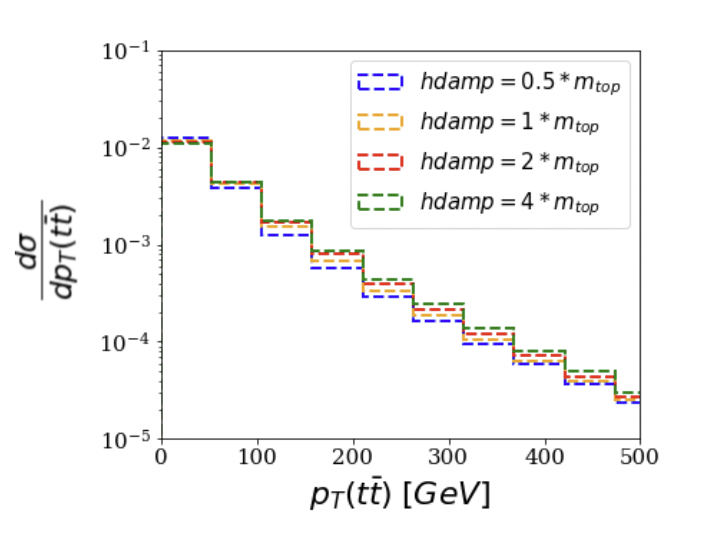}
\includegraphics[width=0.42\textwidth]{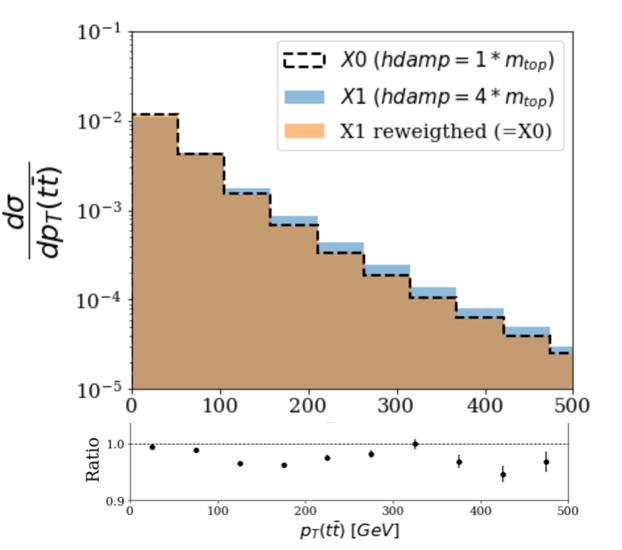}

\caption{Left: The differential cross section as a function of the $\pt$ of the $\ttbar$ system for Powheg HVQ samples generated with four different values of the $\hdamp$ parameter. Right: The differential cross section in $\pt$ of the $\ttbar$ system as a function of the transverse momentum of the $\ttbar$ system for a sample generated with $\hdamp=1\times \mt$ in black and $\hdamp=4\times \mt$ in blue. The orange distribution is the second sample (blue) reweighted to the first one (black).}
\label{fig:pT}
\end{figure}


\section{Results and Conclusions}
The DCTR method was tested by performing the training on samples of four different values of $\hdamp$, respectively equal to ($0.5\times m_{t}$, $1\times \mt$, $2\times \mt$, $4\times \mt$), where $\mt$ corresponds to the default value of CMS (172.5 GeV).
For each value of $\hdamp$, two million events were generated.
The parton-level information is passed as inputs to the DNN. The quadrimomentum and the particle PDGID ($\pt$, $y$, $\phi$, $m$, PDGID) of the $\ttbar$ system and of the additional quark or gluon are presented to the DNN for training, plus the reweighting parameter $\hdamp$. The Particle Flow Network (PFN) [6], which is the composition of two DNNs ($\Phi$ and $F$) is used. The DNN architecture can be seen in Fig.~\ref{fig:NN2}.

\smallskip
The impact of the parameter $\hdamp$ on the $\pt$ of the $\ttbar$ system can be seen on the left plot in Fig.~\ref{fig:pT}. 
To check the performance of the reweighting, the observable  $\pt$ of the $\ttbar$ system was employed.
The weights obtained in the training were applied to reweight a distribution with $\hdamp$ value equal to $4\times \mt$ to the nominal one generated with $\hdamp$ value equal to $1\times \mt$.
The results are shown on the right plot in Fig.~\ref{fig:pT}.
The precision of the reweighting is quantified by the ratio plot between the reweighted sample (orange) and the original one (black).
The original sample and the reweighted one agree within an uncertainty of 5$\%$.
The next step would be to integrate the DCTR method into the CMS software to be used in all future top analyses.

\smallskip

\bibliographystyle{JHEP}
\bibliography{main}
\end{document}